\documentclass[sigconf,screen]{acmart}
\usepackage{microtype}
\usepackage[many]{tcolorbox}
\usepackage{listings}
\usepackage{enumitem}
\setlist[itemize]{leftmargin=*}
\setlist[enumerate]{leftmargin=*}
\usepackage{pifont}
\usepackage{subcaption}
\usepackage{algorithm}
\usepackage[noend]{algpseudocode}
\usepackage{booktabs}
\usepackage{tabularx}
\usepackage{multirow}
\usepackage{makecell}
\usepackage{hyperref}
\usepackage{fvextra}
\input{highlighted-code/default-style.tex}
\usepackage{xcolor}
\usepackage{xspace}
\usepackage{wrapfig}
\usepackage{amsmath,amsthm,amsfonts}
\usepackage[capitalise,nameinlink]{cleveref}
\usepackage{newtxmath} 

\newcommand{\code}[1]{\texttt{\small #1}}

\setcopyright{cc}
\setcctype{by}
\acmDOI{10.1145/3832783.3834440}
\acmYear{2026}
\copyrightyear{2026}
\acmISBN{979-8-4007-2882-2/2026/10}
\acmConference[ASE '26]{Proceedings of the 41st IEEE/ACM International Conference on Automated Software Engineering}{October 12--16, 2026}{Munich, Germany}
\acmBooktitle{Proceedings of the 41st IEEE/ACM International Conference on Automated Software Engineering (ASE '26), October 12--16, 2026, Munich, Germany}
\acmSubmissionID{ase26main-p3621-p}
\received{2026-03-26}
\received[accepted]{2026-06-18}

\begin{document}

\author{Doehyun Baek}
\orcid{0009-0004-0117-1060}
\email{doehyunbaek@gmail.com}
\affiliation{%
  \institution{CISPA Helmholtz Center for Information Security}
  \city{Stuttgart}
  \country{Germany}
}

\author{Michael Pradel}
\orcid{0000-0003-1623-498X}
\email{michael@binaervarianz.de}
\affiliation{%
  \institution{CISPA Helmholtz Center for Information Security}
  \city{Stuttgart}
  \country{Germany}
}

\newcommand\approach{Artisan\xspace}
\newcommand\benchmark{Artisan-Bench\xspace}

\newcommand\goodscript{reproduction script\xspace}
\newcommand\newbug{paper-artifact inconsistencies\xspace}
\newcommand\fastpath{copied-results script\xspace}

\newcommand\staticerror{static error\xspace}
\newcommand\Staticerror{Static error\xspace}
\newcommand\runtimeerror{runtime error\xspace}
\newcommand\Runtimeerror{Runtime error\xspace}
\newcommand\mismatcherror{mismatched results\xspace}
\newcommand\Mismatcherror{Mismatched results\xspace}
\newcommand\copyrepro{copied-results\xspace}
\newcommand\Copyrepro{Copied results\xspace}
\newcommand\downrepro{last-mile reproduction\xspace}
\newcommand\Downrepro{Last-mile reproduction\xspace}
\newcommand\fullrepro{full reproduction\xspace}
\newcommand\Fullrepro{Full reproduction\xspace}

\newcommand\papersetsize{23\xspace}
\newcommand\tablesetsize{60\xspace}
\newcommand\goodscripttsize{44\xspace}
\newcommand\goodscripttsizeoutperform{3.14\xspace}
\newcommand\artisancost{\$0.45\xspace}
\newcommand\artisantime{48 minutes\xspace}
\newcommand\newbugsize{20\xspace}
\newcommand\fastpathsize{30\xspace}

\newcommand\todo[1]{\textcolor{red}{TODO: #1}}

\title{Automated Table Reproduction via Code Generation}

\begin{abstract}
Reproducibility is an important goal in computer science research, e.g., for artifact evaluation and to build upon experimental results of prior work.
Recently, LLM agents are being used to automatically reproduce research results, but they fail to provide executable evidence of reproduction and do not consider the method of reproduction, which limits their usefulness.
We present \approach, an LLM agent that reproduces tables of numeric results, given a paper and its artifact.
The approach is enabled by two key contributions:
First, we frame the reproduction problem as a code generation task, enabling users to audit and re-run the resulting reproduction script independently of the agent.
Second, we design automated judging mechanisms that steer the agent toward correct results without exposing them, while preventing shortcuts like copying precomputed results.
To evaluate \approach, we introduce \benchmark, the first benchmark assessing the ability to generate code that reproduces research results.
\benchmark~comprises 60 tasks derived from 23 software engineering papers.
Our experiments show that \approach~is effective and efficient, with the added benefit of aiding the discovery of 20 new errors in either the paper or artifact.
\end{abstract}



\maketitle

\section{Introduction}
\label{s:introduction}

\emph{Reproducibility}, i.e., the ability of a different team to obtain similar experimental outcomes using the same experimental setup~\cite{acm-artifact-badging-v1_1}, is an important goal in computer science research.
It is crucial for enabling others to verify the results of a paper and build upon them.
Checking reproducibility is a key goal of artifact evaluation~\cite{DBLP:conf/sigsoft/Hermann0S20}, as practiced in software engineering~\cite{DBLP:journals/sigsoft/Krishnamurthi13}, programming languages~\cite{DBLP:journals/cacm/KrishnamurthiV15}, security~\cite{DBLP:conf/ccs/OlszewskiLSWKPU23}, and database~\cite{DBLP:journals/sigmod/AthanassoulisTA22} communities.

Due to its importance, recent work tackles automating the problem of reproducibility with LLM agents.
These approaches can be classified based on the agents' output, summarized in Table~\ref{t:sota_comparison}.
Rating-based agents~\cite{reprobench,heye} output a reproducibility rating.
However, this assessment remains coarse-grained; it lacks granular information on which specific results are reproducible.
Result-based agents~\cite{super, corebench} output the numerical results they have reproduced, offering a fine-grained assessment.
Unfortunately, neither of these approaches provides executable evidence of reproduction, such as a script that can be audited and re-run independently to verify the results.
Hence, when these agents fail to come up with positive outputs, it is unclear whether the error lies with the agent or the artifact.
Even when these agents succeed in producing positive outputs, the method of reproduction, i.e., \emph{how} the results are obtained, is ignored.
That is, the reproduction could simply be the result of guesswork or trivial copying of known results.

Motivated by these limitations, this paper presents \approach, an LLM agent for reproducing research results by generating code that independently reproduces the results.
Given a table with numeric results, along with the corresponding paper and its artifact, the approach generates a \goodscript{} that reproduces the results in the table.
The approach is enabled by two key contributions:
First, we formulate the reproduction problem as a code generation task, where the agent generates a reproduction script that independently reproduces the paper's reported results based on the code and data given in the artifact.
Second, we introduce automated judging mechanisms that steer the agent toward correct results without exposing them, while preventing shortcuts like copying precomputed results.
We focus on the reproduction of tables, as they are the most widely used means of reporting detailed results in research papers.

\begin{table}[t]
\caption{Comparison of existing approaches and~\approach.}
\label{t:sota_comparison}
\footnotesize
\centering
\setlength{\tabcolsep}{3pt}
\begin{tabular}{@{}lccc@{}}
\toprule
 & \shortstack[c]{Rating-based\\agents~\cite{reprobench,heye}}
 & \shortstack[c]{Result-based\\agents~\cite{super, corebench}}
 & \shortstack[c]{\textbf{\approach{}}\\\textbf{(our work)}} \\
\midrule
Fine-grained assessment       &            & \checkmark & \checkmark \\
Executable evidence           &            &            & \checkmark \\
Detects~\fastpath             &            &            & \checkmark \\
\bottomrule
\end{tabular}
\end{table}

We envision our approach to be useful in several scenarios.
First, authors can use \approach to check whether their prepared artifact is sufficiently well-packaged and well-documented to facilitate reproduction.
During our evaluation, we found that good packaging and documentation make the artifacts easier to use not only for people but also for the agents.
Second, evaluators of artifacts can use \approach to automate an important part of the evaluation process, while providing executable evidence of reproduction.
Third, researchers building on prior work can use \approach to quickly reproduce the results of the prior work, which can be time-consuming and error-prone when done manually.
Finally, in case the paper's results disagree with those produced by running an \approach{}-generated \goodscript{}, our approach can support the discovery of \emph{\newbug{}}, defined as the mismatch between the results of the paper and the artifacts.
During our work, we identified \newbugsize{} previously unknown errors in either the papers or artifacts with the help of \approach.

\begin{figure*}[t]
  \centering
  \includegraphics[width=.9\linewidth]{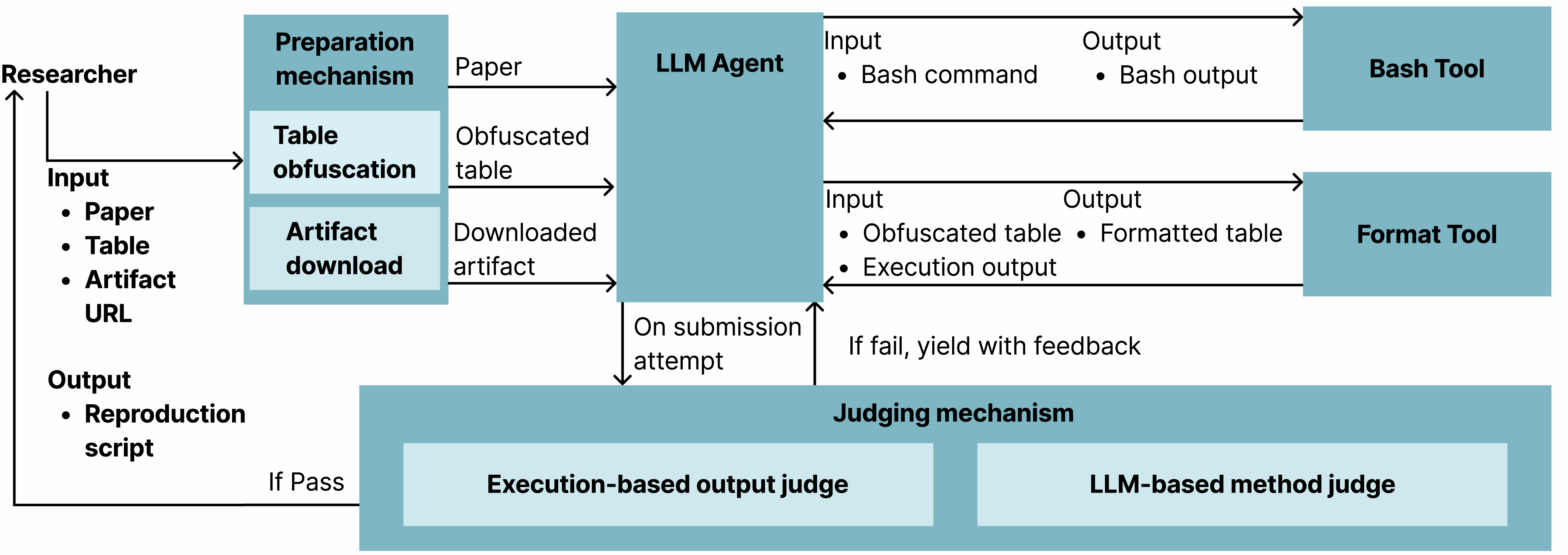}
  \caption{Overview of \approach.}
  \Description{Artisan takes a paper, a table, and an artifact URL. It obfuscates the table, uses an LLM agent with shell and formatting tools to generate a reproduction script, and judges the script's output and reproduction method.}
  \label{f:overview}
\end{figure*}

To evaluate \approach, we present \benchmark, the first benchmark assessing LLM agents' capabilities to generate reproduction scripts.
\benchmark~comprises 60 tasks derived from 23 software engineering papers that cover a diverse range of subfields and techniques, including program analysis, testing, and empirical studies.
To ensure that the 60 tasks are feasible, we manually validate all tasks in \benchmark~for reproducibility.
The benchmark focuses on recent, Dockerized artifacts that deterministically produce results without relying on non-public APIs or special hardware.
This focus ensures that \benchmark is a realistic and relevant benchmark for evaluating the capabilities of LLM agents in reproducing research results, while also being accessible to a wide range of researchers and practitioners.
Evaluating \approach~on the benchmark shows that \approach~is effective, producing \goodscripttsize{}/60 reproduction scripts and outperforming the best available baseline, a vanilla agent (mini-swe-agent), by \goodscripttsizeoutperform{}$\times$ in terms of successful reproduction scripts generated.
The approach is also efficient, costing an average of~\artisancost and~\artisantime{} per task.

In summary, this paper contributes the following:
\begin{itemize}
    \item \approach, an automated research reproduction approach leveraging code generation capabilities of an LLM agent guided by an automated judging mechanism.
    \item \benchmark, the first benchmark that evaluates LLM agents' capabilities in generating reproduction scripts on software engineering papers.
    \item An empirical evaluation showing that Artisan is effective and efficient, while clearly outperforming existing baselines.
    \item Both \approach and \benchmark are publicly available.
\end{itemize}

\setlength{\abovecaptionskip}{1pt}
\setlength{\belowcaptionskip}{4pt}

\section{Approach}
\label{s:approach}

We first define the problem we address (Section~\ref{s:problem_def}) and give a brief overview (Section~\ref{s:overview}).
Then, we present in detail the preparation mechanisms (Section~\ref{s:prep}), tools (Section~\ref{s:tools}), and judging mechanisms (Section~\ref{s:judge}) that enable our approach.

\subsection{Problem Definition}
\label{s:problem_def}

We address the following problem: given a research paper $p$, a table $t$ from $p$, and an artifact URL $u$, automatically generate a reproduction script $s$ that reproduces $t$ using the artifact.
The reproduction script $s$ should contain all logic needed to reproduce the table from scratch, from running the artifact to formatting outputs into the final table.
The reproduction script $s$ may reuse existing artifact code or replace it when necessary.
If the artifact already provides a script that fully regenerates the target table, then $s$ can simply call it.\footnote{As shown in Section~\ref{s:ground_truth}, most artifacts do not include such scripts.}
If reproduction requires multiple scripts, then $s$ should run them in the right order.
If the provided artifact is non-functional (e.g., an obsolete \code{Dockerfile}), $s$ may fix it, so the table can still be reproduced.

Our approach enforces two properties.
First, the produced reproduction script must run without an agent.
Prior work~\cite{super,reprobench,heye,corebench} employs LLM agents to directly output either a rating or results.
Users of these agents must blindly trust the agent output or inspect long agent trajectories.
Generating a reproduction script avoids this, as users can audit and re-run the generated script instead of blindly trusting the agent.
Second, the generated script should reproduce results as fully as possible within the time budget.
Prior work~\cite{super,reprobench,heye,corebench} evaluates whether outputs match, not how they are produced.
In practice, there are multiple ways to reproduce results with different levels of completeness, and ignoring the method of reproduction can misalign with user intent.
For example, if the script merely copies checked-in results rather than actually reproducing them, the user learns nothing about the results' reproducibility.

We focus on the automated reproduction of numerical results in tables for two reasons.
First, tables are the most widely used means to present detailed experimental results in research papers.
Among the 23 papers contained in \benchmark, there are 68 tables with experimental results, outnumbering, e.g., the number of figures and code listings. 
Successful reproduction of the results in the table implies substantial coverage of the experimental results of the paper.
Second, tables are more amenable to automated judging.
While numerical results have well-defined criteria for equality, this is not the case for figures or qualitative results.
This aligns with previous work~\cite{super,reprobench,heye,corebench}, which also focuses on reproducing numerical results.

\subsection{Overview}
\label{s:overview}

\begin{figure*}[t]
  \centering
  \begin{subfigure}[t]{1\textwidth}
    \input{highlighted-code/trajectory-obfuscation.tex}
    \caption{Example trajectory of agent attempting to hardcode expected results using base64 encoding.}
    \label{f:obfuscate_challenge_example}
  \end{subfigure}
  \begin{subfigure}[t]{1\textwidth}
    \input{highlighted-code/trajectory-download.tex}
    \caption{Example trajectory of agent attempting to use wrong artifact repository APIs.}
    \label{f:get_challenge_example}
  \end{subfigure}
  \begin{subfigure}[t]{1\textwidth}
    \input{highlighted-code/trajectory-format.tex}
    \caption{Example trajectory of agent struggling to format execution output into expected format.}
    \label{f:format_challenge_example}
  \end{subfigure}
  \begin{subfigure}[t]{1\textwidth}
    \input{highlighted-code/trajectory-runtime.tex}
    \caption{Example trajectory of agent authoring script that triggers runtime error.}
    \label{f:exec_judge_challenge_example}
  \end{subfigure}
  \begin{subfigure}[t]{1\textwidth}
    \input{highlighted-code/trajectory-copy.tex}
    \caption{Example trajectory of agent authoring script that copies results.}
    \label{f:llm_judge_challenge_example}
  \end{subfigure}
  \caption{Challenges in automated research reproduction.}
  \Description{Five agent trajectories illustrating hardcoded expected results, an incorrect artifact download, unformatted output, a runtime error, and copied checked-in results.}
  \label{f:challenges}
\end{figure*}

Figure~\ref{f:overview} outlines \approach.
\approach is a ReAct-style~\cite{yao2022react} LLM agent, i.e., an LLM-based system that autonomously uses tools to achieve a goal~\cite{DBLP:conf/icse/BouzeniaDP25}.
Before entering the agentic loop, \approach prepares the inputs via table obfuscation and artifact download (Section~\ref{s:prep}).
The agent uses two tools: a general \emph{bash tool} and a task-specific \emph{format tool} (Section~\ref{s:tools}).
Once the agent submits a script, a two-tier judge is invoked to verify the correctness of both the output and method (Section~\ref{s:judge}).
In case the submitted script fails either of the two judging steps, the agent is given corresponding feedback and continues its iteration.
Otherwise, \approach~outputs the submitted \goodscript.

\subsection{Preparation Mechanisms}
\label{s:prep}

Before entering the agentic loop, \approach~performs two preparation steps: table obfuscation and artifact download.

\subsubsection{Table Obfuscation}
\label{s:obfuscate}

The key objective of our approach is guiding the agent toward the expected results.
One option is to provide the expected table as a reference for the agent to follow.
However, during the development of \approach, we found that LLM agents may directly use the expected table when given in unobfuscated form.
Figure~\ref{f:obfuscate_challenge_example}~shows an example trajectory of an agent exploiting the given expected table, where the agent encodes the expected table using base64 encoding and decodes it at runtime to avoid directly hard-coding the expected table in the reproduction script.

To address this challenge, our preparation mechanism obfuscates the numerical values of the expected table before giving it to the agent.
Figure~\ref{f:obfuscate_example} shows an example of the obfuscation mechanism.
Given the input table (Figure~\ref{f:obfuscate_input}), our obfuscation mechanism replaces the digits that appear at the ``A+'' column (annotations) and ``TW'' column (total warnings) with question marks, resulting in the obfuscated table (Figure~\ref{f:format_input_result}).
The obfuscated table is then given to the agent as the table to reproduce, guiding the agent without revealing the expected outputs.

\begin{figure*}[t]
  \centering
  \begin{minipage}[t]{0.48\textwidth}
    \centering
  \begin{subfigure}[t]{0.48\linewidth}
    \input{highlighted-code/obfuscation-original.tex}
    \caption{Table before obfuscation.}
    \label{f:obfuscate_input}
  \end{subfigure}\hfill
  \begin{subfigure}[t]{0.48\linewidth}
    \input{highlighted-code/obfuscation-masked.tex}
    \caption{Table after obfuscation.}
    \label{f:obfuscate_output}
    \label{f:format_input_result}

  \end{subfigure}
  \begin{subfigure}[t]{0.98\linewidth}
    \input{highlighted-code/execution-log.tex}
    \caption{Execution output.}
    \label{f:format_input_log}
  \end{subfigure}
    \caption{Example obfuscation and execution outputs for Table~3 of ScType~\cite{sctype}.}
    \Description{Three subfigures show the original table, the table with numeric values replaced by question marks, and the corresponding execution log.}
    \label{f:obfuscate_example}
    \label{f:format_input}
  \end{minipage}
  \hfill
  \begin{minipage}[t]{0.48\textwidth}
    \centering
  \begin{subfigure}[t]{0.98\linewidth}
    \input{highlighted-code/results-expected.tex}
    \caption{Part of the expected results.}
    \label{f:judge_execution_example_1}
  \end{subfigure}
  \begin{subfigure}[t]{0.98\linewidth}
    \input{highlighted-code/results-actual.tex}
    \caption{Part of the actual results.}
    \label{f:judge_execution_example_2}
  \end{subfigure}
  \begin{subfigure}[t]{0.98\linewidth}
    \input{highlighted-code/results-feedback.tex}
    \caption{Feedback given to the agent.}
    \label{f:judge_execution_example_3}
  \end{subfigure}
    \caption{Example of the mismatched results feedback for Table~2 of Drosos et al.~\cite{bloat}.}
    \Description{Three subfigures show the expected results, the actual results with a different unresolved-call count, and feedback that masks the mismatched count with question marks.}
    \label{f:judge_execution_example}
  \end{minipage}
\end{figure*}

\subsubsection{Artifact Download}
\label{s:get}

A typical artifact URL points to a repository hosting the artifact, such as Zenodo, Figshare, or GitHub.
Given such a URL, the agent must first download the artifact file before it can proceed with reproducing the results.
However, we found that agents often struggle to find the correct way to download the actual artifact file given the artifact URL.
Figure~\ref{f:get_challenge_example}~shows an example, where the agent is given the artifact's Zenodo URL and presumes the artifact is downloadable through a canonical file-download endpoint.
However, this endpoint does not exist, and the agent downloads an HTML file containing ``Page not found''.
The agent, assuming a successful download, attempts to unzip the file and realizes it is not a valid ZIP archive.

To address this challenge, we provide a download mechanism, which the agent can invoke with only the URL of the research artifact.
Given this URL, the download mechanism first queries the repository to obtain the list of files contained in the artifact.
It then downloads each file, caching any file that has not been retrieved before.
Finally, it locates the README file by searching for files whose names begin with “README” among all downloaded contents, including those stored inside ZIP or TAR archives.
Although simple, this strategy effectively saves the agent several steps that would otherwise be spent reasoning about repository-specific APIs.
The download mechanism currently supports Zenodo, Figshare, and GitHub.

\subsection{Tools}
\label{s:tools}

In the agentic loop, \approach can invoke tools to interact with the environment.
We provide a standard bash tool and a custom format tool.
The rationale for using these tools is that the bash tool offers general CLI interaction, which we found to be sufficient for most of the reproduction tasks, while the format tool addresses a specific challenge that we observed during development.

\subsubsection{Bash Tool}
\label{s:bash}

We utilize the bash tool of mini-swe-agent~\cite{minisweagent}~to handle generic CLI usage tasks, encompassing file reading, file writing, and running commands.
The vanilla mini-swe-agent we compare against in the evaluation has only the bash tool.

\subsubsection{Format Tool}
\label{s:format}

Even after an agent successfully reproduces results by running a sequence of commands, the outputs rarely match the exact tabular format shown in the paper.\footnote{We use markdown tables as the expected format, as LLM agents are proficient at generating them.}
While the agent may try to write a script to format the output into a table, we find that it commonly struggles with this task.
Figure~\ref{f:format_challenge_example} illustrates a typical trajectory: the agent runs the right commands and submits the~\goodscript~containing them, but the execution output is formatted differently from the expected table, and hence our judge (Section~\ref{s:judge_execution}) rejects it.
The agent inspects the output yet does not realize that the formatting discrepancy is the cause, and repeatedly reruns the same command without making progress.

To address this challenge, we implement the format tool: an LLM-based formatter that takes an obfuscated table plus arbitrary execution outputs and produces properly formatted results.
Execution outputs may be unstructured logs or tables in other formats (e.g., CSV or LaTeX).
Figure~\ref{f:format_input} shows an example: the execution output is a log file without tabular structure (Figure~\ref{f:format_input_log}), which does not match the obfuscated table (Figure~\ref{f:format_input_result}).
The format tool prompts an LLM to fill in question marks in the obfuscated table with digits extracted from the execution output, and the LLM locates the annotation count (6) and warning count (1) for MarginSwap and populates the table.
To mitigate LLM non-determinism, we cache format-tool results so identical inputs return cached outputs.

\subsection{Judging Mechanism}
\label{s:judge}

A key component of \approach is to automatically judge whether the reproduction script generates correct output with an appropriate method.
We present a two-tier judging mechanism that combines an execution-based output judge and an LLM-based method judge.

\subsubsection{Execution-Based Output Judge}
\label{s:judge_execution}

Previous work on rating-based agents~\cite{reprobench,heye} and output-based agents~\cite{super, corebench} for reproducibility asks agents to yield simple numerical results as the final output.
However, as~\approach~asks the agents to output a~\goodscript, there are additional failure modes that arise when the script triggers syntax or runtime errors.
Often, the agent successfully executes commands to reproduce the research results, but there is some fault in the submitted script, which can often be addressed by giving an error message as feedback.
Figure~\ref{f:exec_judge_challenge_example}~shows an example trajectory of an agent facing this challenge.
The agent writes a \goodscript~that utilizes awk to extract research results from an execution output, but uses invalid awk syntax.
Accepting this script would not be helpful, as the user would have to debug the script to find the syntax error, which is time-consuming and requires expertise in bash scripting and awk.

To address this challenge, the output judge runs the submitted script in a fresh container environment and classifies the result into four categories:
\begin{itemize}
  \item \textbf{\Staticerror}: The submitted script is missing from the expected path or fails to run because of a syntax error.
  \item \textbf{\Runtimeerror}: The submitted script runs but produces a runtime error.
  \item \textbf{\Mismatcherror}: The submitted script runs to completion without any error, but the results differ from the expected values.
  \item \textbf{Success}: The submitted script runs to completion without any error, and the results match the expected values.
\end{itemize}

In the first three cases, \approach gives corresponding feedback to the agent, allowing it to continue its iteration.
Importantly, in the case of \mismatcherror{}, the output judge reveals parts of the result that the agent got right, but not the parts it got wrong.
Figure~\ref{f:judge_execution_example}~gives an example of this feedback mechanism.
Suppose Figure~\ref{f:judge_execution_example_1}~is the expected result of the reproduction, and Figure~\ref{f:judge_execution_example_2}~is the actual result obtained after running the submitted \goodscript.
The output judge detects that the ``Resolved Count'' matches (7,799,929) but the ``Unresolved Count'' does not match (260,249 expected vs. 168,482 actual).
Hence, it provides the feedback shown in Figure~\ref{f:judge_execution_example_3}, which further guides the agents without leaking the expected results.

\subsubsection{LLM-Based Method Judge}
\label{s:judge_llm}
\begin{figure}[t]
  \centering
  \begin{subfigure}[t]{0.46\textwidth}
    \input{highlighted-code/method-copied.tex}
    \caption{Example~\copyrepro{}~script for Table~2 of Unimocg~\cite{DBLP:conf/se/HelmRKRM25}.}
    \label{f:judge_llm_example_1}
  \end{subfigure}
  \begin{subfigure}[t]{0.46\textwidth}
    \input{highlighted-code/method-last-mile.tex}
    \caption{Example~\downrepro{}~script for Table~3 of Wallner et al.~\cite{pmsat}.}
    \label{f:judge_llm_example_2}
  \end{subfigure}
  \begin{subfigure}[t]{0.46\textwidth}
    \input{highlighted-code/method-full.tex}
    \caption{Example~\fullrepro{}~script for Table~2 of Wallner et al.~\cite{pmsat}.}
    \label{f:judge_llm_example_3}
  \end{subfigure}
  \caption{Examples of three kinds of reproduction methods.}
  \Description{Three shell scripts respectively copy checked-in results, process existing raw data, and rerun an experiment before processing its results.}
  \label{f:judge_llm_example}
\end{figure}

While developing~\approach, we encountered many cases where the LLM agent produces a \goodscript that yields the correct results for the wrong reasons.
A typical scenario is when the agent hard-codes the expected results inside the reproduction script or simply copies them from precomputed results that are part of the artifact.
Figure~\ref{f:llm_judge_challenge_example}~shows an example trajectory of an LLM agent facing this challenge.
Upon discovering that the artifact contains the research results of Table~1 checked in, the agent writes a~\goodscript~that simply uses these results and signals completion.
As the overall goal is to reproduce the research results, such ``shortcuts'' are undesirable.

To address this challenge, once the submitted script passes the output judge, \approach invokes the method judge, an LLM-as-judge evaluating the method of reproduction.
The method judge classifies the reproduction method into three categories:
\begin{itemize}
  \item \textbf{\Copyrepro{}}: The submitted script just copies the expected values from the checked-in results or the paper without any meaningful reproduction.
  An example is shown in Figure~\ref{f:judge_llm_example_1}, where the script copies checked-in results.
  \item \textbf{\Downrepro{}}: The submitted script performs a lightweight reproduction, skipping some time-consuming steps.
  An example is shown in Figure~\ref{f:judge_llm_example_2}, where the script uses checked-in raw data and processes it to generate results.
  Although this falls short of full reproduction, we judge this to be a valid reproduction.
  This is because the full reproduction of some experiments may take hundreds of hours (as in the example), and verifying that the analysis of the raw data matches the paper's results is still valuable.
  \item \textbf{\Fullrepro{}}: The submitted script performs the most complete reproduction possible.
  This is the ideal case, especially if it can be achieved within reasonable time.
  An example is shown in Figure~\ref{f:judge_llm_example_3}, where the script performs the full reproduction, from raw data generation to analysis, on one example.
\end{itemize}

If the method judge classifies the submitted script as either \downrepro{} or \fullrepro{}, \approach accepts the submitted script as the final~\goodscript.
In contrast, scripts classified as \copyrepro{} are rejected, and the approach yields control back to the agent, asking it to find an alternative, more complete reproduction method.

\section{Benchmark}
\label{s:benchmark}

The following presents \benchmark, the first benchmark for evaluating LLM agents' capabilities in generating reproduction scripts for software engineering research papers.

\subsection{Scope}
\label{s:scope}

We focus the scope of the benchmark on recent, Dockerized artifacts that deterministically produce results without relying on non-public or commercial APIs, or on special hardware.
Focusing on Dockerized artifacts is motivated by the fact that Docker and related virtualization technologies are now standard in artifact evaluation~\cite{icse2024-ae,fse2024-ae,ase2024-ae}.
Focusing on deterministic results ensures that our benchmark itself is deterministic, allowing us and others to reliably evaluate and compare approaches on the benchmark, without the results being confounded by non-determinism in the artifacts.
The motivation for excluding artifacts using non-public APIs and specialized hardware (e.g. GPUs) is to maintain affordability and portability of the benchmark.
Taken together, these restrictions ensure that the benchmark is realistic, reliable, and accessible to a wide range of researchers and practitioners, while still being challenging for state-of-the-art LLM agents.

\subsection{Manual Validation}
\label{s:requirements}

To ensure that all tasks in \benchmark are feasible, we manually validate the reproducibility of each task before including it in the benchmark.
This step is motivated by SWE-Bench~\cite{jimenez2024swebench}, where subsequent efforts, such as SWE-Bench Verified~\cite{chowdhury2024swebenchverified} and Wang et al.~\cite{DBLP:journals/corr/abs-2503-15223}, highlight the importance of manual validation to avoid overestimating or underestimating agent capabilities.
Thus, we attempt a manual reproduction of all papers with an eight-hour time budget per paper.
The manual reproduction attempt is performed by the first author, who has multiple years of research experience in software engineering and is familiar with reproducibility and artifact evaluation.
The manual reproduction attempt either results in a ground truth reproduction script, or it fails if the time budget runs out.

\subsection{Paper Selection}
\label{s:paper_select}

Guided by the above scope and requirements, we curate a set of \papersetsize~papers following the process below.
\begin{enumerate}
  \item \textbf{Recent papers with badges.} We gather 114 papers and artifacts from four software engineering conferences (ICSE, FSE, ASE, ISSTA) in 2024 with ACM \emph{Artifacts Available} and \emph{Artifacts Evaluated-Reusable} badges~\cite{acm-artifact-badging-v1_1}. We do not consider \emph{Results Validated-Results Reproduced} and \emph{Results Validated-Results Replicated} badges as no 2024 ICSE, FSE, ASE, ISSTA papers have received these badges yet.
  \item \textbf{Docker.} We exclude 42 papers whose artifacts are not packaged using Docker, which yields 72 remaining papers.
    To implement this filter, we case-insensitively search Markdown and PDF files for the term ``docker'' and exclude any artifacts that do not mention the term ``docker''.
  \item \textbf{Non-public APIs.} We exclude nine papers that rely on non-public APIs, yielding 63 remaining papers.
    First, the first author inspected artifacts to find keywords that signal the use of non-public APIs.
    As a consequence, we search Python files and Jupyter notebooks for ``openai'' or ``anthropic''.
    We also search Markdown files for ``etherscan''.
    This excludes seven papers using OpenAI APIs and two papers using Etherscan APIs.
  \item \textbf{Specialized hardware.} We exclude 19 papers that require GPUs or other specialized hardware, yielding 44 remaining papers.
    To support this filter, we search for common GPU keywords (e.g., \code{cuda}, \code{cudnn}, \code{tensorflow}, \code{nvidia-smi}, \code{gpus}) in source code files.
  \item \textbf{Reproducibility budget.} We exclude 21 papers to enforce our eight-hour reproducibility budget, yielding the final set of \papersetsize~papers.
    We first remove 12 papers via a search for ``hour'', as they report runtimes above the limit (four $>$12 hours, seven $>$24 hours, one $>$125 hours).
    We then attempt to manually reproduce the remaining 32 papers and exclude nine where no table is reproducible within the budget.
    Causes for exclusion are missing datasets (2x), raw data (2x), and generation scripts (2x), non-deterministic outputs (2x), and one inaccessible Docker image (1x).
\end{enumerate}

The selected \papersetsize~papers cover a wide range of software engineering areas.
Specifically, we classify five papers as empirical studies, five as static analysis, two as security, two as dynamic analysis, two as software maintenance, one paper in each of fuzzing, mutation testing, model-based testing, mobile applications, program repair, test automation, and defect prediction.

\subsection{Task Selection}
\label{s:task_select}

From the \papersetsize~selected papers, we curate \tablesetsize~tasks.
Starting from all 89 tables contained in the \papersetsize~papers, we exclude 21 non-result tables, which contain descriptions of evaluation setups (14x) or the approach (7x).
We also exclude six tables that we fail to reproduce during our paper selection process, one table that requires 400 GB of RAM to reproduce (Table 3 of Unimocg~\cite{DBLP:conf/se/HelmRKRM25}), and one table that contains only non-deterministic results (Table 3 of PPT4J~\cite{DBLP:conf/icse/Pan00Z0024}).

When trying to reproduce tables, it may become obvious that either the paper or artifact contains some errors, leading to inconsistencies between the two.
As part of this work, we found~\newbugsize{} cases where either the paper or the artifact is erroneous, which we discuss in detail in Section~\ref{s:inconsistencies}.
In these cases, plus three cases where the paper authors claim the artifact's results could be slightly different from those in the paper~\cite{DBLP:journals/pacmse/Song0LCR24}, we use the corrected version of the table as the task in our benchmark.
Moreover, in some cases, we include only a part of the table as the task:
for 15 tables because we could only partially reproduce the results, and for four tables because some results are non-deterministic and hence excluded.

\subsection{Ground Truth Scripts}
\label{s:ground_truth}

For each task, we prepare a ground truth script that reproduces the table from scratch using the artifact.
In total, we create 30 \fullrepro scripts and 30 \downrepro scripts.
Note that we do not utilize these ground truth scripts for our automated judging mechanism (Section~\ref{s:judge}); rather, they serve as executable evidence that reproduction is possible for each table in the evaluation set.

To quantitatively characterize the complexity of the ground truth scripts, we report that the number of non-blank, non-comment LoC per script is 71.8 on average, with a median of 57.
We also manually analyze all the ground truth scripts for very simple scripts and find six scripts where the expected table can be produced with a single command.
In contrast, most scripts require multiple commands, complex formatting logic, or sometimes even fixes of outdated or buggy artifact code.

\subsection{Development Process}
\label{s:development_process}

We developed \approach and \benchmark in tandem.
Preliminary evaluation results informed our design choices when developing \approach.
For example, when we discovered that LLM agents submit copied-results scripts without the actual reproduction, we devised the method judge to mitigate the problem.

\section{Evaluation}
\label{s:evaluation}

We address the following research questions:

\begin{itemize}[labelindent=\parindent,leftmargin=*]
\item \textbf{RQ1. Effectiveness}:
How effective is \approach{} at generating successful reproduction scripts?
\item \textbf{RQ2. Efficiency}:
How efficient is \approach{} in terms of monetary cost, token consumption, and execution time?
\item \textbf{RQ3. Ablation study}:
What is the impact of our technical contributions, i.e., output judge, method judge, and format tool?
\item \textbf{RQ4. Copied-results scripts}:
How do agents construct \copyrepro{} scripts, and how accurate is the method judge at detecting them?
\item \textbf{RQ5. Paper-artifact inconsistencies}:
What \newbug{} does \approach{} help discover?
\end{itemize}

\subsection{Experimental Setup}

\subsubsection{Agent Framework, LLMs, and Settings}

\begin{table*}[t]
  \small
  \centering
  \caption{Comparison of \approach{} with baselines. Success and failure numbers are sums over tasks. Cost and time are averages.}
  \label{t:baseline}
  {\setlength{\tabcolsep}{3pt}
\begin{tabular}{llccccccccc}
\toprule
\multirow{2}{*}{} & \multirow{2}{*}{} & \multicolumn{2}{c}{Success} & \multicolumn{4}{c}{Failure} & \multirow{2}{*}{\shortstack{Cost}} & \multirow{2}{*}{\shortstack{Tokens}} & \multirow{2}{*}{\shortstack{Time}} \\
\cmidrule(lr){3-4} \cmidrule(lr){5-8}
 & & Full reprod. & Last-mile reprod. & Copied-result & Mismatch & Runtime err. & Static err. & & & \\
\midrule
SWE-agent & w/ DeepSeek-3.2-Reas. & 0 & 1 & 0 & 6 & 7 & 46 & \$0.03 & 575.8k & 8.9 min \\
& w/ GPT-5-mini & 0 & 0 & 9 & 10 & 10 & 31 & \$0.05 & 363.9k & 3.1 min \\
& w/ GPT-5.1 & 0 & 3 & 4 & 15 & 8 & 30 & \$0.32 & 513.0k & 6.5 min \\
OpenHands & w/ DeepSeek-3.2-Reas. & 3 & 0 & 0 & 6 & 6 & 45 & \$0.03 & 796.4k & 12.8 min \\
& w/ GPT-5-mini & 0 & 1 & 1 & 25 & 15 & 18 & \$0.04 & 473.3k & 23.1 min \\
& w/ GPT-5.1 & 5 & 7 & 4 & 12 & 5 & 27 & \$0.24 & 577.1k & 21.6 min \\
mini-swe-agent & w/ DeepSeek-3.2-Reas. & 3 & 2 & 5 & 16 & 3 & 31 & \$0.03 & 437.2k & 50.4 min \\
& w/ GPT-5-mini & 0 & 0 & 4 & 39 & 13 & 4 & \$0.06 & 344.8k & 33.1 min \\
& w/ GPT-5.1 & 9 & 5 & 12 & 22 & 9 & 3 & \$0.26 & 328.2k & 13.8 min \\
\midrule
Artisan & w/ DeepSeek-3.2-Reas. & 11 & 8 & 4 & 15 & 5 & 17 & \$0.05 & 565.9k & 63.1 min \\
& w/ GPT-5-mini & 15 & 9 & 6 & 24 & 6 & 0 & \$0.14 & 699.2k & 42.9 min \\
& w/ GPT-5.1 & 24 & 20 & 2 & 10 & 4 & 0 & \$0.45 & 479.0k & 48.0 min \\
\midrule
Ablations & \shortstack{w/o Output Judge} & 13 & 14 & 1 & 19 & 13 & 0 & \$0.28 & 272.2k & 12.3 min \\
& \shortstack{w/o Method Judge} & 9 & 18 & 21 & 8 & 3 & 1 & \$0.43 & 405.3k & 32.0 min \\
& \shortstack{w/o Format Tool} & 17 & 18 & 9 & 13 & 2 & 1 & \$0.46 & 499.6k & 52.0 min \\
\bottomrule
\end{tabular}
}

\end{table*}

We implement \approach{} based on mini-swe-agent~\cite{minisweagent} (version 1.14.4).
The preparation mechanisms, judging mechanisms, and the format tool are implemented in Python~3.12.10.
For all baselines and \approach{}, we experiment with two proprietary models (GPT-5.1-2025-11-13 and GPT-5-mini-2025-08-07) and one open-weight model (DeepSeek V3.2 Reasoner).
As the LLM behind the format tool and method judge, we use GPT-5-mini-2025-08-07.
To limit resource consumption per task, we set a step limit of 30, a cost limit of USD~1, and a time limit of eight hours.

\subsubsection{Baselines}

As baselines, we select mini-swe-agent~\cite{minisweagent}, SWE-agent~\cite{sweagent}, and OpenHands~\cite{openhands}.
We select mini-swe-agent, because \approach{} builds on it, enabling a thorough empirical study of \approach{}'s contributions.
We choose SWE-agent and OpenHands as established state-of-the-art approaches for general software engineering tasks.
All baseline agents are equipped with the same user prompt, which explains the task and suggests a workflow for addressing it.
They are given access to the same information as \approach{}, except for the obfuscated table and the judging mechanisms.
The reason for not evaluating against existing rating-based agents~\cite{heye,reprobench} and result-based agents~\cite{super, corebench} is twofold.
First, they formulate the task differently, making it difficult to adapt them to our setting without substantial modifications.
Second, the main contributions of those papers are benchmarks of reproduction tasks in other domains, whereas their agent designs are mostly task-specific adaptations and prompt improvements, which do not generalize to software engineering artifacts.

\subsubsection{Metrics}

To evaluate effectiveness (RQs~1 and~3), we classify the scripts submitted by the baselines and \approach{} into six categories, following the classification introduced in Section~\ref{s:judge}.
For the three failure categories, we automatically classify them based on the execution outputs.
For the three success categories, i.e., where the results of the submitted script match the expected result, we manually classify the reproduction method.
The manual classification is primarily performed by the first author, with repeated discussions with another, senior author to ensure consistency and to resolve ambiguous cases.
As a single metric of overall effectiveness, we report the sum of full reproduction scripts and last-mile reproduction scripts generated by each approach, which we consider successful reproduction.

\subsection{RQ1: Effectiveness}
\label{s:effectiveness}

\subsubsection{End-to-end Effectiveness}

The left-hand side of Table~\ref{t:baseline} shows the effectiveness of \approach and the baselines on \benchmark.
\approach~with GPT-5.1 gives the best result, successfully reproducing the expected table in~\goodscripttsize/60 tasks.
Notably, \approach outperforms all baselines even when using weaker LLMs than the baselines, which we attribute to two key factors.
First, \approach reduces the number of scripts with static or runtime errors.
With GPT-5.1, it produces only 4/60 static or runtime-error scripts (6/60 with GPT-5-mini), versus 38/60 for SWE-agent and 32/60 for OpenHands.
Second, \approach creates many more scripts with correct output and method.
\approach with GPT-5.1 lowers all failure types compared to using a weaker model like DeepSeek, unlike mini-swe-agent which mainly shifts errors to \mismatcherror or \copyrepro.

\subsubsection{Reasons for Success}


To analyze the reasons for successful reproduction, we select tasks where all of the approaches succeed in reproduction (excluding SWE-agent, as it rarely succeeds) and tasks where only one approach succeeds.
We find a pattern for success on 5/6 tasks where all OpenHands, mini-swe-agent and \approach successfully reproduce the expected table: the instructions for reproducibility are clearly documented in the README.
If this condition is met, the reproduction task becomes significantly easier, as the agent can effortlessly locate commands to exercise the artifact.
Figure~\ref{f:readme_table} shows a part of the README that provides instructions to reproduce Table~3 of Drosos et al.~\cite{bloat}.
In our user prompt given to the agents, we instruct them to run grep commands to search for the expected table, which returns a hit to this part of the README.
Although these five tasks might not be very challenging for an agent, we argue that they reveal a desirable characteristic that artifacts should display.

\begin{figure}
  \begin{minipage}{1\linewidth}
    \begin{minipage}{\linewidth}
      \small
      \input{highlighted-code/readme-example.tex}
      \captionof{figure}{Part of the README of Drosos et al.~\cite{bloat} artifact.}
      \label{f:readme_table}
    \end{minipage}

    \begin{minipage}{\linewidth}
      \input{highlighted-code/r-example.tex}
      \captionof{figure}{Part of the R code of Zid et al. \cite{pythonic} artifact. Comments are ours.}
      \label{f:pythonic_mixup}
    \end{minipage}
  \end{minipage}
  \hfill
  \Description{Two code excerpts: README instructions for reproducing a table and R code showing p-value adjustment applied after values were copied.}
\vspace{-1em}
\end{figure}


Not all tasks in \benchmark are so straightforward.
For 24 tasks, only \approach produces a successful reproduction.\footnote{SWE-agent and OpenHands do not have any tasks where they are the sole successful approach. mini-swe-agent is the only successful approach on one task.}
We present an analysis of Table 5 of Zid et al. \cite{pythonic}, where \approach is the only agent capable of reproduction.
What makes this particularly interesting is that Table~3 of the same paper is a much easier table to reproduce, and is also reproduced by mini-swe-agent.
In the paper, Tables~3 and~5 report BH-adjusted p-values.\footnote{BH refers to the Benjamini-Hochberg procedure for controlling the false discovery rate.}
In contrast, the artifact code for Table~5 records raw p-values and applies BH only afterward, which means the recorded outputs remain unadjusted (Figure~\ref{f:pythonic_mixup}).
Trying to reproduce Table 5, \approach first writes a reproduction script with the raw p-value and submits it.
However, our output judge rejects this submission and notifies the agent that the p-values do not match.
Then, after probing the artifact, the agent realizes that the paper uses the adjusted p-value, unlike the R code in the artifact.
The agent adds code to post-process the raw p-value with BH adjustment, and passes the automated judging.
This showcases that our automated judging mechanism is especially crucial in challenging cases where the reproduction is not straightforward.

\subsubsection{Reasons for Failure}


Across the 10 tasks that no approach (including \approach) can reproduce, we observe three failure modes.
(1) Irreproducible checked-in results (5/10): For Tables 2, 4, and 5 of Bouzenia and Pradel~\cite{action}, results used in the paper are checked-in, but re-running the notebook yields different results.
Agents try to reproduce the checked-in results, repeatedly fail judging, and hit the step limit.
Tables 2 and 3 of NPETest~\cite{npetest} fail for similar reasons.
This suggests a need for stronger guidance when there are conflicting results.
(2) Step limit reached (4/10): For Table~2 of Eder et al.~\cite{interference}, Table~2 of Li et al.~\cite{rust}, Table~3 of Bouzenia et al.~\cite{dypybench}, Table~4 of ProveNFix~\cite{DBLP:journals/pacmse/Song0LCR24}, \approach with GPT-5.1 hits the step limit before successfully reproducing the table.
Manual analysis of the four trajectories reveals that 55\% of the steps are spent on exploration (grep, cat, and sed commands), without actually exercising the artifact.
This motivates using a higher step limit to reproduce more complex artifacts or having better guidance that aids the reproduction in fewer steps.
(3) Buggy artifact code (1/10): For Table~1 of AXA~\cite{axa}, a Docker-shipped bash script used for counting changes in lines of code contains errors.
The erroneous bash script contains a stray ``+'' and a missing term in a sum, leading to different results.
Agents execute the erroneous script, obtain incorrect counts, and fail to repair it.
This motivates an automated program repair technique that is applicable to the artifacts to aid the reproduction process.

\subsection{RQ2: Efficiency}
\label{s:efficiency}

\paragraph{Money}
We report the average per-task monetary cost of LLM queries on the right-hand side of Table~\ref{t:baseline}.
For \approach, this includes the costs of the format tool (Section~\ref{s:format}) and the method judge (Section~\ref{s:judge_llm}).
We generally observe \approach~to impose higher costs than the baselines, which is caused by it taking all the steps required to actually run the experiments instead of failing early.
That said, with \$0.45 per task on average with GPT-5.1, the approach is still practical and cost-efficient compared to human effort.
For \approach with GPT-5.1, the median cost is \$0.388 (standard deviation: \$0.227; 90th percentile: \$0.744; maximum: \$1.001).
Broken down by outcome, the median costs are \$0.322 for \fullrepro, \$0.372 for \downrepro, \$0.292 for \copyrepro, \$0.669 for \mismatcherror, and \$0.783 for \runtimeerror.
Even with a cheaper model, e.g. DeepSeek V3.2 Reasoner, \approach remains quite effective (19/60 successful reproductions) but is highly cost-efficient (\$0.05 per task).
This shows that \approach with cheaper models can outperform baselines with more expensive models both in terms of effectiveness and efficiency.

\paragraph{Token Consumption}
We report the average per-task token consumption of LLM queries on the right-hand side of Table~\ref{t:baseline}.
We find that the use of stronger models leads to generally lower token usage.
We suspect this is due to stronger models being able to solve the tasks with similar difficulties with fewer tokens.

\begin{figure}[t]
  \centering
  \includegraphics[width=\linewidth]{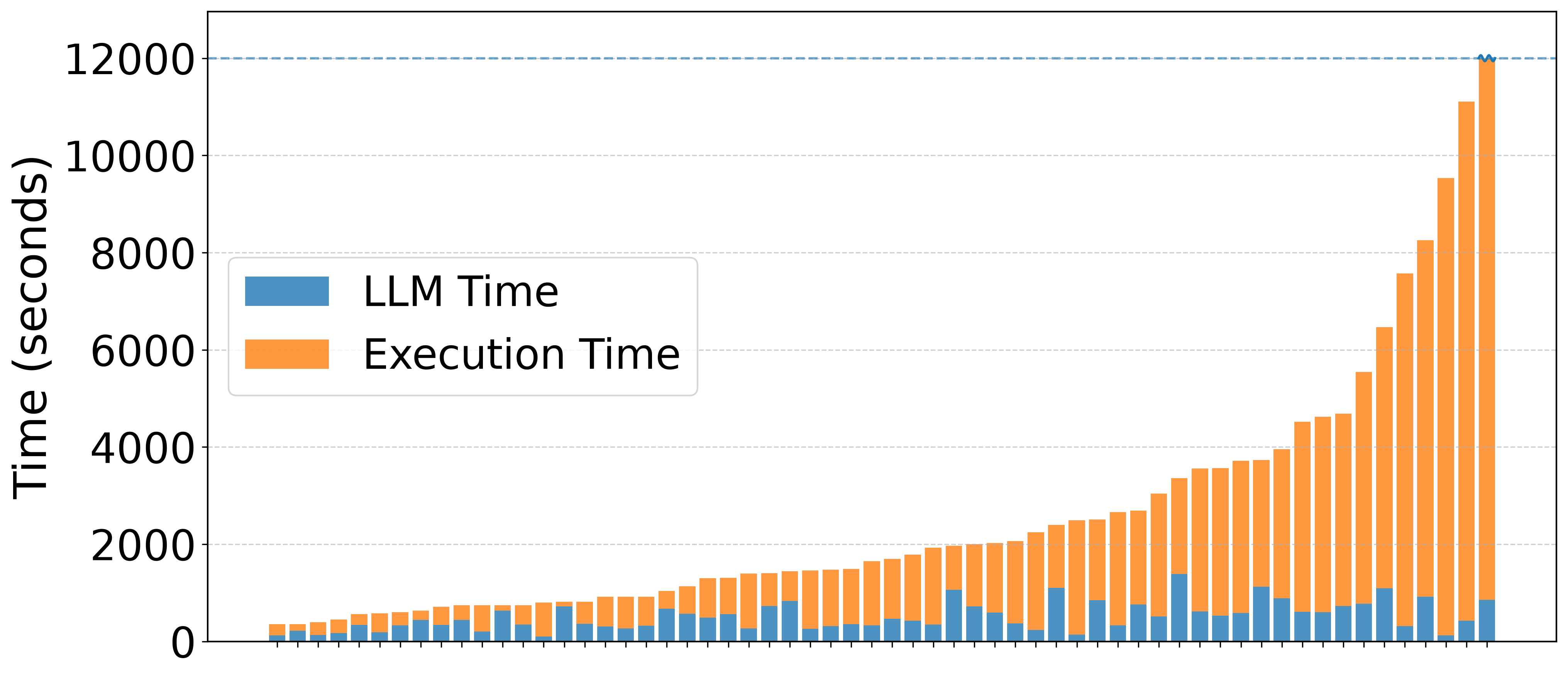}
  \caption{Breakdown of wall-clock time by activity.}
  \Description{Stacked bars compare execution and LLM time across run-duration buckets; execution accounts for most time, especially for longer runs.}
  \label{f:time_breakdown}
\end{figure}

\paragraph{Time}
We report the average wall-clock time spent by each approach per task on the right-hand side of Table~\ref{t:baseline}.
The time measurements include all agent activities, including the time to run the reproduced experiments themselves.
Thus, \approach taking more time than the baselines is a sign that it successfully performs long-running experiments.
Figure~\ref{f:time_breakdown} shows where time is spent during a task for \approach with GPT-5.1, where \emph{Execution time} refers to
running agent commands and the judging mechanism.
Execution time dominates the overall time for long runs, while LLM time contributes noticeable overhead for shorter runs, confirming our observation that taking longer time per task is due to it successfully running long-running experiments.
One might suspect that higher execution time could be a sign of repeated failures or inefficiencies.
However, average time spent across all evaluated successful runs (n=215) is 2,038 seconds, whereas the average across all failed runs (n=685) is 1,588 seconds, which is substantially less.
This shows that repeated failure is not a major reason for long execution time.
For \approach with GPT-5.1 ($n=60$), the median wall-clock time is 27.9 minutes (standard deviation: 69.22 minutes; 90th percentile: 94.0 minutes; maximum: 487.6 minutes).
Broken down by outcome, the median wall-clock times are 20.4 minutes for \fullrepro, 29.0 minutes for \downrepro, 14.5 minutes for \copyrepro, 50.4 minutes for \mismatcherror, and 70.2 minutes for \runtimeerror.

\subsection{RQ3: Ablation Study}
\label{s:ablation}


The lower part of Table~\ref{t:baseline} summarizes our ablation study, which uses three variants of~\approach that exclude the output judge, the method judge, and the format tool, respectively.
Omitting the output judge increases the number of reproduction scripts with \runtimeerror (4 to 13) and \mismatcherror (10 to 19).
In principle, agents can check for errors or mismatches themselves by running the generated scripts, but enforcing this check on submission attempt makes a significant difference.
Reduction in monetary cost and wall-clock time is not a desirable sign; it merely indicates that the agent exits early with non-functional reproduction scripts.
Without the method judge, we see a dramatic increase in \copyrepro scripts (2 to 21), showing that the method judge is crucial for inducing the agent to seek a more complete reproduction method.
Omitting the format tool decreases successful reproductions (44 to 35) and increases \copyrepro scripts (2 to 9).
These two effects are related: When given the liberty to format the reproduced table as they like, the agent often directly copies the expected results.
Providing the format tool to the agent reduces such behavior by simplifying the task of correctly formatting the output.

\subsection{RQ4: Copied-Results Scripts}
\label{s:llmtools}

\begin{table}
  \small
  \centering
  \caption{Accuracy of method judge.}
  \label{t:confusion}
  \begin{tabular}{lccc}
  \toprule
  & \multicolumn{3}{c}{\textbf{Predicted}} \\
  \cmidrule(lr){2-4}
  \textbf{Actual} & \textbf{Full} & \textbf{Last} & \textbf{Copy} \\
  \midrule
  \textbf{Full}      & 28 & 1  & 1  \\
  \textbf{Last-Mile} & 11 & 18 & 1  \\
  \textbf{Copy}      & 0  & 3  & 27 \\
  \bottomrule
\end{tabular}

\end{table}

\subsubsection{How Do Agents Construct \copyrepro Scripts?}
\label{s:copyrerpo}


While developing and evaluating \approach, we observed many cases where the agent generated copied-results scripts (Section~\ref{s:judge_llm}).
While similar misalignment issues are reported in other domains~\cite{DBLP:journals/corr/abs-2503-11926, macdiarmid2025naturalemergentmisalignmentreward}, to our knowledge, this is the first report of such misalignment in automated research reproduction.
To understand how agents construct \copyrepro scripts, we sample \fastpathsize cases and manually classify their strategies:
In 12/\fastpathsize cases, agents directly copy checked-in research results.
In 7/\fastpathsize cases, agents convert the paper PDF to text and copy expected tables.
In other 7/\fastpathsize cases, agents copy outputs from included Jupyter notebooks.
In 3/\fastpathsize cases, agents infer results from surrounding paper text despite table obfuscation.
Finally, in one case, an agent downloads the preprint by searching for it on arXiv.
These diverse strategies indicate that preventing specific misalignment types is infeasible, confirming our decision to use a method judge to detect \copyrepro~scripts.

\subsubsection{Accuracy in Detecting Copied-Results Scripts}
\label{s:eval_judge}

\begin{figure}
  \small
  \input{highlighted-code/method-judge-example.tex}
  \caption{Part of the \fullrepro scripts reproducing Table~3 Zid et al. \cite{pythonic}, which is misclassified as \copyrepro scripts.}
  \Description{A shell script starts a Docker container, runs an R analysis, and copies the generated table to the reproduction output.}
  \label{f:full_to_copy}
\end{figure}


We evaluate the method judge using 30 sampled \copyrepro scripts and 60 ground truth scripts (Section~\ref{s:ground_truth}).
Table~\ref{t:confusion} presents the accuracy results.
The method judge achieves 81\% accuracy (73/90).
Successful reproductions misclassified as \copyrepro occur in only 3\% of cases (2/60), which is acceptable as \approach can still progress toward more complete scripts.
Conversely, 10\% of \copyrepro scripts are misclassified as successful reproductions (3/30), a rate mitigated by manual validation of final reproduction scripts.

Figure~\ref{f:full_to_copy} shows an example where the method judge misclassifies a \fullrepro script as a \copyrepro script.
Here, \code{Table-3-RQ1-lambda.txt} is generated by running an R script, so the output copied with \code{cat} is not a checked-in result.
However, the method judge incorrectly identifies this file as checked-in and flags it as a \copyrepro script.
Since the method judge cannot execute scripts, it cannot determine if the copied file is precomputed.
Using a more capable method judge, e.g., a separate agent, could reduce such misclassifications.

\subsection{RQ5: Paper-Artifact Inconsistencies}
\label{s:inconsistencies}

\begin{table}[t]
  \small
  \centering
  \setlength{\tabcolsep}{2pt}
  \caption{Inconsistencies found between paper and artifact. TI stands for table index. \#DE stands for the count of different entries for a table. AD stands for artifact difference. TE stands for transcription error. RE stands for rounding error. AE stands for artifact error. \textsuperscript{\S} denotes confirmation by the authors.}
  \label{t:inconsistencies}
  \begin{tabular}{l r r r r r}
\toprule
\textbf{Paper} & \textbf{TI} & \textbf{\#DE} & \textbf{Paper Value} & \textbf{Artifact Value} &\textbf{Reason} \\
\midrule
action \cite{action} & 1 & 4 & 0.30 & 0.32 &AD\textsuperscript{\S} \\
action \cite{action} & 2 & 29 & 33.8 & 36.2 &AD\textsuperscript{\S} \\
action \cite{action} & 3 & 1 & 66.3 & 66.4 & RE \\
action \cite{action} & 4 & 21 & -55.14 & -62.63 &AD\textsuperscript{\S} \\
action \cite{action} & 5 & 5 & -10.41 & -99.78 &AD\textsuperscript{\S} \\
axa \cite{axa} & 1 & 2 & 14 & 1 & AE\textsuperscript{\S}\textsuperscript{1} \\
bazel \cite{bazel} & 5 & 4 & 28 & 29 & TE\textsuperscript{\S} \\
dypybench \cite{dypybench} & 3 & 1 & 1701 & 1543 & AE\textsuperscript{\S}\textsuperscript{2} \\
interference \cite{interference} & 2 & 4 & $8.13 \times 10^{285960}$ & $8.58 \times 10^{506}$ & TE\textsuperscript{\S} \\
lasapp \cite{lasapp} & 1 & 1 & 1549 & 1553 &AD\textsuperscript{\S} \\
llm \cite{llm} & 3 & 4 & 1.15 & 0.15 & TE\textsuperscript{\S} \\
npetest \cite{npetest} & 2 & 27 & 100 & 68 & AD\textsuperscript{\S} \\
npetest \cite{npetest} & 3 & 2 & 73 & 74 & AD\textsuperscript{\S} \\
pmsat \cite{pmsat} & 3 & 1 & 0.6 & 0.5 & RE \\
pmsat \cite{pmsat} & 5 & 1 & 17.33 & 17.3 & RE \\
pythonic \cite{pythonic} & 3 & 5 & 0.14 & 0.24 & TE \\
pythonic \cite{pythonic} & 9 & 1 & 0.04 & 0.05 & RE \\
rust \cite{rust} & 1 & 6 & 98.88\% & 98.93\% &AD\textsuperscript{\S}\\
sctype \cite{sctype} & 3 & 2 & 1 & 2 & TE \\
urcrat \cite{urcrat} & 1 & 1 & 72199 & 72201 &AD\textsuperscript{\S}\\
\bottomrule
\end{tabular}

\scriptsize \textsuperscript{1}Fixed by https://zenodo.org/records/18292903

\scriptsize \textsuperscript{2}Fixed by https://github.com/sola-st/DyPyBench/commit/bccfda7

\end{table}

By using \approach, we encounter \newbugsize~inconsistencies between papers and their artifacts.
We identify these cases by noticing inconsistencies between the paper results and the output of the~\goodscript~when the \goodscript{} appears to contain no apparent error.
We do not include cases where artifacts themselves warn about possible inconsistencies; thus, all \newbugsize~cases we report are newly discovered inconsistencies that have not been disclosed publicly before.
Table~\ref{t:inconsistencies} summarizes the \newbug~we find.
Overall, we find \newbugsize~\newbug~from 13 papers, i.e. 57\% of the studied papers.
All these 13 papers underwent an artifact evaluation in major SE conferences and were awarded the \emph{Artifacts Evaluated -- Reusable} badge.
Still, with the aid of \approach{}, we find inconsistencies in up to 29 entries in a single table, with values differing by up to $8.13 \times 10^{285960}$.

To provide more insight into the \newbug we find, we manually analyze the reasons for the inconsistencies, leading to five categories:
\begin{itemize}
  \item \textbf{Artifact difference} (9/\newbugsize): The artifact (code and data) used for the paper differs from the artifact publicly available.
  \item \textbf{Transcription error} (5/\newbugsize): Errors in manually transcribed data.
  \item \textbf{Rounding error} (4/\newbugsize): Inconsistencies due to wrong rounding.
  \item \textbf{Artifact error} (2/\newbugsize): Errors in the artifact. In one case, the code contained an error, and in the other case, the data contained an error.
\end{itemize}

We report our findings (excluding rounding errors, as they are obvious but minor issues) to the paper authors, and 14/16 of the reported inconsistencies are confirmed.
For the two artifact errors, both have already been fixed by the authors.
For the three paper errors (Table~2 of Eder et al.~\cite{interference}, Table~3 of Nam et al.~\cite{llm}, and Table 5 of Alfadel and McIntosh~\cite{bazel}), the authors communicated plans to update the errors.
Our approach enables early detection of such reproducibility issues, saving time and avoiding their propagation into future research that builds on the affected papers.

The discovery of \newbug is supported by \approach and the reproduction scripts it generates, but it is not entirely automated.
Given the reproduction script, it is ultimately up to a human expert to decide if there is indeed a bug, and if so, where the bug lies.

\section{Discussion}
\label{s:discussion}

\subsection{Limitations}
\label{s:limitations}

While \approach is effective at reproducing research results, it has several limitations.
First, the approach currently only supports textual results in tables.
Addressing this limitation would require the development of new judging mechanisms that can evaluate the correctness of non-textual results, e.g., arbitrary figures, which is a non-trivial task.
Second, \approach expects the outputs to exactly match the expected results and thus cannot reproduce non-deterministic results.
Thus, \approach could not be applied to itself, as \approach makes use of an LLM, which outputs non-deterministic results.
Future work could relax this requirement by comparing results with a certain tolerance, e.g., for performance metrics, and by enforcing multiple executions.
Finally, the current agent runtime and output judge support Debian-based Docker images only, which could be addressed through additional engineering effort, which we avoided as most artifacts support Linux and Docker.

\subsection{Threats to Validity}
\label{s:threats}

Our selection criteria for papers and tasks in \benchmark may introduce bias, as we focus on recent, Dockerized artifacts that deterministically produce results without relying on non-public APIs or special hardware.
This focus is motivated by the desire to allow us and others to evaluate the capabilities of LLM agents without requiring monetary resources (e.g., for API keys) while minimizing confounding factors such as software evolution and hardware availability.
The downside of this focus is that the benchmark may underrepresent the challenges of reproducing results from older, not well-packaged artifacts, artifacts that rely on commercial APIs, and research that relies on very long-running experiments.
We plan to research how to expand the benchmark and \approach to cover such artifacts in the future.
Constructing a ground truth script for each task in \benchmark and evaluating the reproduction method of the evaluated approaches involves manual effort, which may introduce bias.
Due to the high manual effort, this process is primarily conducted by the first author.
However, to mitigate potential bias, a second, senior author has reviewed the ground truth scripts, and all authors have discussed ambiguous cases.
Our results are influenced by the step limit (30 steps) and cost limit (\$1), which we impose to keep the evaluation cost and time reasonable.
With GPT-5.1, step limits are hit by SWE-agent (18/60), OpenHands (22/60), mini-swe-agent (6/60), and Artisan (16/60), suggesting that early stopping in most tasks limits effectiveness more than incomplete exploration.


\section{Related Work}
\label{sec:relatetd}

\paragraph{Reproducibility of Research Results}

Reproducibility of research results is a broad issue in general science~\cite{Baker20161500SL}.
For computer science, various efforts like artifact evaluation processes~\cite{DBLP:conf/sigsoft/Hermann0S20}, reproducibility program~\cite{DBLP:journals/jmlr/PineauVSLBdFL21}, and empirical evaluation on reproducibility of LLM performance~\cite{DBLP:journals/corr/abs-2510-25506} aim to enhance the reproducibility of published research results.
Our work contributes to the effort of enhancing reproducibility by automating part of the reproduction process.

\paragraph{Automated Research Reproduction}

Recent work leverages LLM agents for reproducibility, defined as enabling a different team to obtain similar experimental outcomes using the same setup~\cite{acm-artifact-badging-v1_1}.
SUPER~\cite{super} introduces a benchmark evaluating agents' ability to set up and execute tasks from research repositories.
CORE-Bench~\cite{corebench} also evaluates agents on reproducing research results, expanding to more scientific disciplines (e.g., social science, medicine), while REPRO-Bench~\cite{reprobench} targets social science papers.
Unlike SUPER and CORE-Bench, REPRO-Bench is rating-based: the agent outputs a reproducibility score from 1 to 4. 
Heye et al.~\cite{heye} develop an LLM-based system that supports artifact evaluation by outputting a reproducibility score, setting up environments, and identifying pitfalls.
Our approach and benchmark differ from prior work as follows:
(1) We formulate the problem as a code generation task, enabling independent assessment of generated reproduction scripts.
(2) We provide an automated judging mechanism that evaluates both the output and the method of reproduction.

\paragraph{Automated Research Replication}

There is a related but distinct problem of research replication, defined as the ability of a different team to obtain similar experimental outcomes using a different setup~\cite{acm-artifact-badging-v1_1}.
Liang et al.~\cite{DBLP:journals/pacmse/LiangBBDFF024} studied LLMs for replicating empirical software engineering research, finding them to be effective at identifying assumptions and boilerplate code.
PaperBench~\cite{DBLP:conf/icml/StaraceJSACMDMK25} evaluates LLM agents on replicating ML papers and reports a top replication score of 21.0\%, below the human baseline.
PaperCoder~\cite{DBLP:journals/corr/abs-2504-17192} is a multi-agent LLM framework for automated research replication via planning, analysis, and generation.
Xiang et al.~\cite{DBLP:journals/corr/abs-2504-00255} introduce SciReplicate-Bench, a replication benchmark for 36 NLP papers and a multi-agent approach for it.
Our work fundamentally differs in the problem addressed (reproduction vs.\ replication).

\paragraph{General Software Engineering LLM Agents}

There is a growing trend of using LLM agents for software engineering tasks~\cite{cacm2026-agents}.
Such agents cover a wide range of tasks, including automated program repair~\cite{DBLP:conf/icse/BouzeniaDP25}, issue solving~\cite{Zhang2024a,sweagent}, engineering tasks in machine learning~\cite{Chan2025}, and setting up and running test suites~\cite{issta2025_ExecutionAgent}.
The task of reproducing research results partially overlaps with software engineering tasks, but has distinct challenges, such as validating the method of reproduction.
Our work contributes a novel agent and a novel benchmark for this task.

\section{Conclusion}
\label{sec:conclusion}

We present \approach, an automated research reproduction approach leveraging code generation capabilities of an LLM agent guided by an automated judging mechanism.
To evaluate \approach, we develop \benchmark, a new benchmark that evaluates LLM agents' ability in generating reproduction scripts on software engineering papers.
Our evaluation shows that \approach~is effective and efficient, having the added benefit of aiding the discovery of \newbug.

\section*{Data-Availability Statement}
The artifact for this paper is available at \cite{artisan-artifact}.
Most recent code is available at \url{https://github.com/doehyunbaek/artisan}

\section*{Acknowledgments}
This work was supported by the German Research Foundation (DFG; projects 492507603, 516334526, and 526259073).

\bibliographystyle{ACM-Reference-Format}
\bibliography{references,referencesMichael}

\end{document}